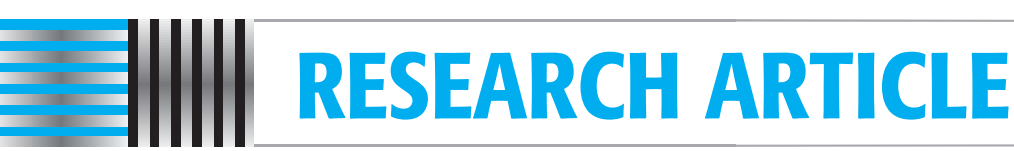

# Data-Efficient Excavation Force Estimation for Wheel Loaders

**ARMIN ABDOLMOHAMMADI, NAVID MOJAHED, SHIMA NAZARI, AND BAHRAM RAVANI**

Department of Mechanical and Aerospace Engineering, University of California at Davis, Davis, CA 95616, USA

Corresponding author: Armin Abdolmohammadi (abdolmohammadi@ucdavis.edu)

This work was supported in part by Komatsu Ltd.

**ABSTRACT** Accurate prediction of excavation forces is critical for enabling autonomous operation and optimizing control strategies in earthmoving machinery. Conventional approaches often depend on extensive data collection or computationally expensive simulations across multiple soil types, which limits their scalability and adaptability. This study presents a data-efficient framework that calibrates soil parameters using force measurements from the preceding bucket-loading cycle. The proposed method is based on an analytical soil-tool interaction model formulated through the fundamental earthmoving equation, and employs a multi-stage optimization procedure during the loading phase to identify relevant soil parameters. These estimated parameters are then used to predict excavation forces in the subsequent cycle, allowing the system to adapt its control inputs without relying on large-scale datasets or machine learning model training. The framework is validated through high-fidelity simulations in the Algoryx Dynamics engine under different soil types and excavation trajectories, achieving root-mean-square prediction errors between 10% and 15%. This cycle-to-cycle adaptation demonstrates strong potential for scalable, online force estimation and efficient path planning in wheel loader operations.

## I. INTRODUCTION

The autonomous operation of earthmoving machinery is a prominent focus in modern engineering and research due to advances in robotics, automation, and control alongside the growing labor shortage in the construction industry [1]. Wheel loaders are one form of autonomous machinery widely used for transporting and dumping materials in high-frequency loading operations across diverse environments. The bucket loading phase accounts for peak energy demand and poses significant challenges due to soil variability and unpredictable tool-material interactions [2]. Optimizing energy use during this phase can significantly improve the efficiency of wheel loader operations.

To enable frameworks that can plan and control bucket motion in soil, it is essential to accurately estimate the forces acting on the bucket during excavation. Force estimation methods generally fall into two categories: *reactive* and *predictive*. Reactive methods estimate forces in real time from sensor data, such as hydraulic pressure and joint motion, using inverse dynamics, observers, or data-driven models. These models include purely data-driven approaches, like a Radial Basis Function (RBF) neural network trained on real-time sensor data [3], and hybrid methods that combine Gaussian Process Regression with rigid body dynamics and a disturbance Kalman filter [4]. Model-based techniques using inverse dynamics and observer design have also been proposed. However, these models offer greater stability and precision at the cost of increased system modeling complexity [5], [6], which is effective for feedback control but not suitable for task planning.

Predictive models estimate resistive forces from geometry, soil, and motion, enabling path and task planning prior to execution. This modeling can be performed numerically, such as through Finite Element Analysis (FEA) or the Discrete Element Method (DEM), primarily in simulations [7], [8], or analytically for faster estimation that is effective for task planning and control. Analytical models directly estimate excavation resistance forces by incorporating physical mechanisms, such as soil wedge theory, which assumes the formation of a triangular failure zone ahead of the cutting tool. These models relate resistive forces to soil properties and tool geometry (e.g., rake angle, tool depth) [9].

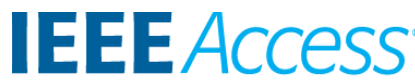

A range of analytical formulations have been developed to capture these resistive interactions with varying levels of physical detail and computational complexity. The Swick and Perumpral model [10] incorporates interface friction, rake angle effects, and variable soil depth. The Gill and Vanden Berg formulation [11] accounts for tool geometry, speed, and soil deformation. Hemami [12] extends analytical modeling to robotic manipulators by coupling soil resistance with dynamic arm control, while the Hettiaratchi and Reece model [13] refines soil resistance estimation by decomposing cutting forces into passive earth pressure and friction components, which is particularly suited to variable-depth scenarios.

Another widely adopted force model is the Fundamental Earthmoving Equation (FEE), originally proposed by Reece [14] and later simplified by McKyes [15]. The FEE is widely used in force prediction models for its simplicity, analytical tractability, and minimal input requirements compared to other methods, making it attractive for control, parameter identification, and embedded implementation. The model has been applied in various excavation contexts, including automated digging control [16], real-time soil parameter estimation [17], [18], hybrid neural-physics resistance models [19], reinforcement learning-based controllers [20], and multibody simulations of wheel loaders [21]. A comprehensive review of these and related models is provided in [9], and [22].

All of these analytical force prediction models rely on soil parameters, such as cohesion, internal friction angle, adhesion, and sinkage characteristics. These parameters are typically obtained through parameter identification methods, such as field probe tests (e.g., Cone Penetration Test (CPT), Pressuremeter Test (PMT), and Dilatometer Test (DMT) [23]) or controlled laboratory experiments [24]. Since such procedures are impractical in construction environments, on-machine estimation methods have emerged that allow for inferring parameters during operation using vehicle-mounted sensors.

Among the on-machine parameter identification methods, learning-based approaches use prior data to train models that classify material types or regress soil properties. Examples include proprioceptive force signal classification [25] and physics-informed neural networks that infer properties from kinematic and control data [26]. While effective, these methods typically require large, labeled datasets and substantial offline training.

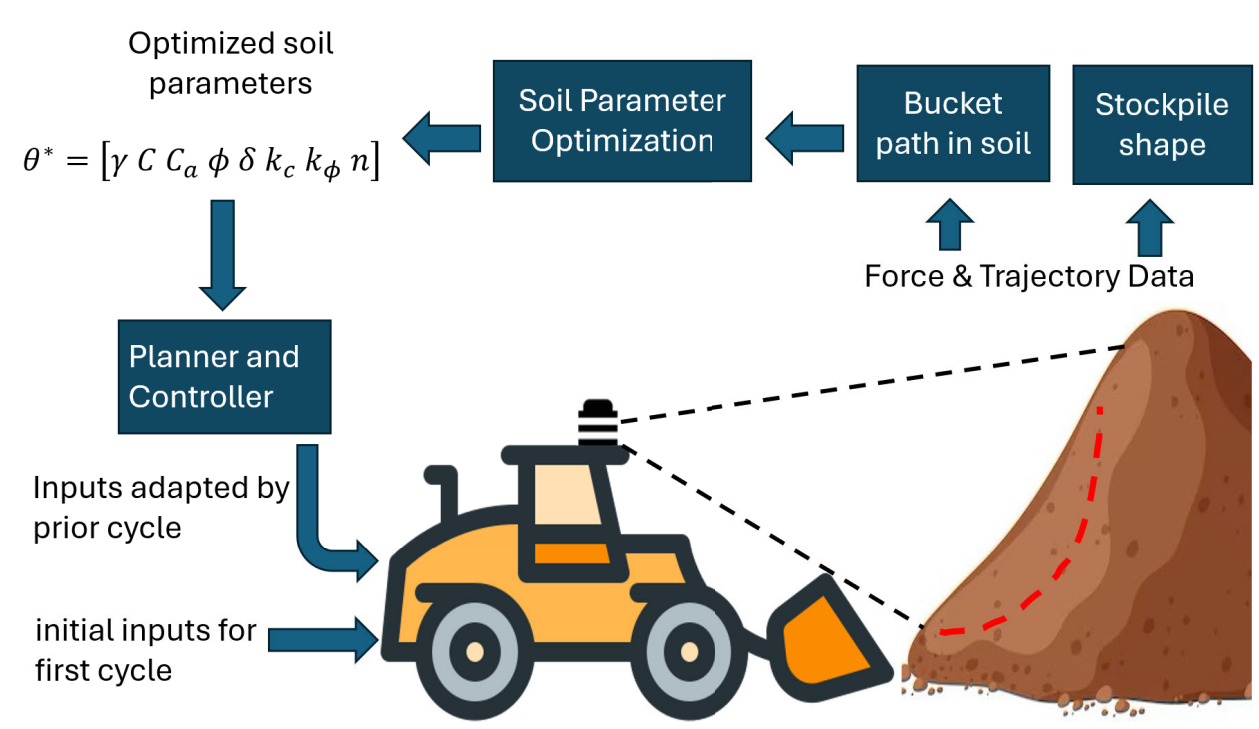

**FIGURE 1.** **Overview of the proposed framework: A predefined digging path is executed, and bucket trajectory along with stockpile geometry is recorded for soil parameter optimization. The estimated parameters inform planning and control for the next excavation cycle.**

Non-learning-based methods, on the other hand, directly estimate parameters using model-based optimization with real-time force measurements. Tan et al. [18] applied a Newton-Raphson method to estimate soil density and internal friction angle, which was later extended to include additional parameters by Althoefer et al. [27]. Zhao et al. [28] further refined the model through the integration of fuzzy logic with an enhanced FEE model for simultaneous parameter and force prediction. However, these methods either restrict the number of estimated parameters, assuming some values are known, or attempt to estimate all parameters at once, resulting in high-dimensional optimization that is prone to convergence issues and computational inefficiency.

### *A. GAPS AND CONTRIBUTIONS*

Three gaps emerge in the literature. First (1), learning-based approaches require extensive labeled datasets or simulations and often rely on aggregated parameters over multiple excavation cycles [19], [26]. Second (2), non-learning-based models suffer from high-dimensional optimization that hinders convergence. Third (3), FEE-based models exhibit force discontinuities due to trigonometric singularities, while existing fixes, such as [29], have introduced added computational cost and terrain sensitivity.

This work presents a data-efficient, cycle-based parameter optimization framework that is designed to run within the short interval between consecutive digging cycles, thereby enabling timely adaptation of operations to changing soil conditions. To address gap (1), a non-learning-based approach is adopted, employing a modified version of the FEE model based on [30] that accounts for sloped terrain and incorporates Bekker's load-sinkage theory [31] for soil compaction. To address gap (2), a multi-stage optimization strategy is proposed that decomposes the parameter fitting task into sequential sub-problems, thereby reducing parameter coupling and improving convergence and computational efficiency. Finally, to address gap (3), continuity constraints derived from trigonometric relationships in failure surface geometry are embedded directly into the minimization procedure used to identify the failure surface, thereby eliminating force discontinuities.

**TABLE 1.** List of symbols, parameters, and variables used in this study.

| Symbol | Description | Unit |
|---|---|---|
| | **Soil Parameters** | |
| $\gamma$ | Soil density | kg/m$^3$ |
| $C$ | Soil cohesion | N/m$^2$ |
| $C_a$ | Adhesion coefficient between soil and bucket | N/m$^2$ |
| $\phi$ | Internal friction angle | rad |
| $\delta$ | External friction angle | rad |
| $k_c$ | Cohesive modulus of deformation | N/m$^{n+1}$ |
| $k_\varphi$ | Frictional modulus of deformation | N/m$^{n+2}$ |
| $n$ | Exponent of soil deformation | – |
| | **FEE dimensionless coefficients** | |
| $N_\gamma$ | Bearing capacity factor for soil unit weight | – |
| $N_c$ | Bearing capacity factor for cohesion | – |
| $N_a$ | Bearing capacity factor for adhesion | – |
| $N_q$ | Bearing capacity factor for surcharge or overburden | – |
| | **Environment Parameters** | |
| $\alpha$ | Stockpile slope angle | rad |
| $\beta$ | Failure wedge angle | rad |
| | **Loader Parameters** | |
| $w$ | Bucket width | m |
| $b$ | Bucket thickness | m |
| $W_b$ | Bucket weight | kg |
| | **Forces and Pressures** | |
| $F_T$ | Tangential cutting force on bucket | N |
| $F_N$ | Normal force on bucket | N |
| $F_R$ | Resultant excavation force on bucket | N |
| $P$ | Penetration pressure (Bekker load–sinkage model) | Pa |
| | **Parameter Sets** | |
| $\theta$ | Full soil parameter vector, $[\gamma, C, C_a, \phi, \delta, k_c, k_\varphi, n]^T$ | – |
| $\theta_1$ | Stage 1 parameter set, $[C_a, \delta, k_c, k_\varphi, n]^T$ | – |
| $\theta_2$ | Stage 2 parameter set, $[\gamma, C, \phi]^T$ | – |
| $\theta_3$ | Stage 3 parameter set, $[k_c, k_\varphi, n]^T$ | – |

This framework is designed to be integrated into a complete excavation planning and control system, allowing a wheel loader to autonomously refine its perception of the environment cycle-by-cycle. As illustrated in Fig. 1, the full process begins by capturing the initial geometry of the soil pile using the loader's assumed onboard perception system. A preliminary digging cycle is then executed, during which the bucket's trajectory is tracked via a kinematic model and excavation forces are measured using load cells embedded in the loader's hinge joints. The collected trajectory, force data, and soil pile geometry are used to optimize soil parameters in an analytical FEE model. These parameters are then used to predict resistive forces and inform the excavation strategy for the next cycle. By repeating this process after each digging cycle, the system continuously refines its understanding of the soil. This paper focuses on excavation force estimation of this full cycle.

Section II of this paper presents the methodology used in this research, including the analytical soil model and bucket-soil interaction and force formulation. Section III describes the proposed multi-stage optimization of the soil parameter. Section IV showcases the results obtained from the implementation of the method. Section V presents key takeaways and implications for future research.

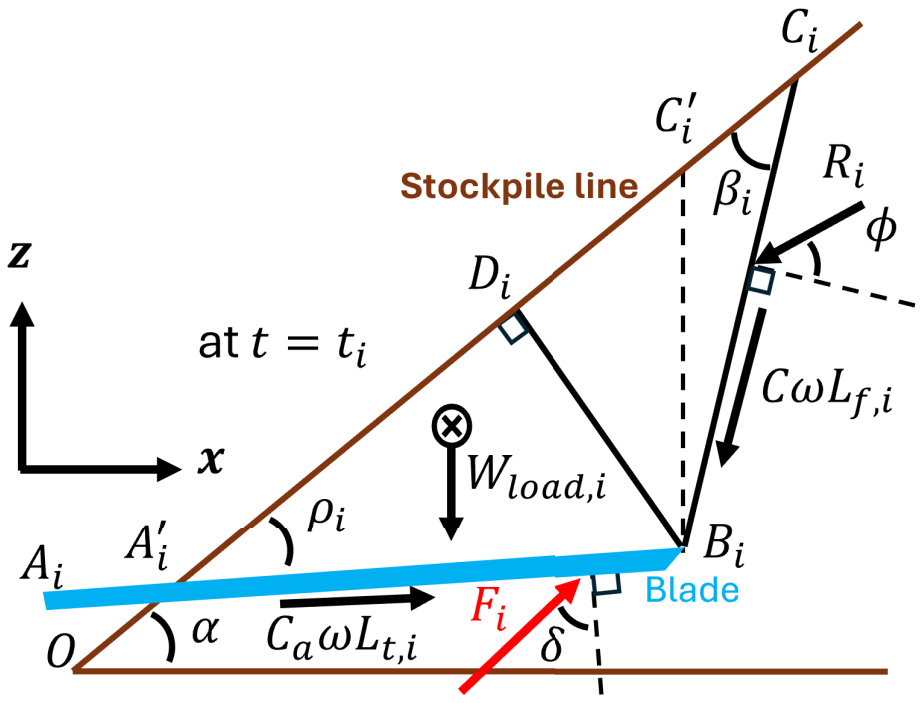

**FIGURE 2.** Free body diagram of the soil wedge acting on the blade segment $A'B$, representing a snapshot of the blade interacting with the soil at a specific time instant. All parameters that vary with time and are captured at $t = t_i$ are denoted with the subscript $i$.

## II. METHODOLOGY

This section presents the framework for modeling soil-tool interaction during excavation. The foundation of this framework is the modified version of the FEE [15], which was chosen for its balance between physical accuracy and computational efficiency [32]. The section then details the incorporation of Bekker's load-sinkage formulation [31] to account for soil compaction effects as well as the trigonometric derivation used to enforce continuity in the soil failure surface geometry.

### A. FUNDAMENTAL EARTHMOVING EQUATION (FEE)

FEE provides a physics-based framework for estimating soil resistive forces during cutting and excavation. Originally introduced by Reece in 1964 [14], the model applies soil mechanic principles to describe forces acting on a cutting blade penetrating soil. McKyes later simplified this framework using the trial wedges method, assuming a straight failure surface and uniform surcharge pressure to estimate forces on tillage equipment [15]. The FEE has since been extended to sloped terrain and applied to structural design and load analysis of excavators [33] and wheel loaders [30]. More recently, Yao used the modified FEE in a model predictive framework to optimize bucket-loading trajectories [32]. This model captures the full set of resistive forces acting on a soil wedge as the blade penetrates the terrain.

Assuming the blade penetrates the soil, the excavation path is discretized into $N$ points at times $t = t_i$, where $i \in \{1, \ldots, N\}$. The free body diagram of the soil wedge at each point is shown in Fig. 2. The stockpile surface is represented by the line $OC$, inclined at angle $\alpha$, and the blade is modeled as segment $AB$, with length $L_{t,i}$ and angle $\rho_i$ relative to the stockpile. The tip of the blade is point $B$, and the failure surface, represented by line $BC$, is assumed to be straight, consistent with McKyes' wedge theory [15]. The forces acting on the soil wedge include the net frictional and normal force $F_i$ between the blade and the wedge, acting at the external friction angle $\delta$, the force $R_i$ from the undisturbed soil at the internal friction angle $\phi$, the cohesion force $CwL_{f,i}$ along the failure surface, and the adhesion force $C_awL_{t,i}$ along the blade. Both $L_{t,i}$ and $L_{f,i}$ vary over time as the blade advances. All soil, environment, and loader parameters are listed in Table 1.

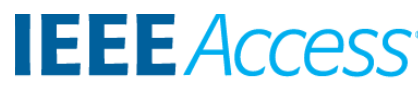

$C_awL_{t,i}$ and $CwL_{f,i}$ can be derived from the geometry of the blade and the soil. $W_{load,i}$ is estimated during operation using the methods described in [3], and [5]. The forces $F_i$ and $R_i$ are unknown. To compute $F_i$, equilibrium conditions are assumed, and Newton's force-balance equations in the $x$ and $z$ directions are applied. Using these equations, $F_i$ can be expressed in terms of $W_{load,i}$, $C_awL_{t,i}$, and $CwL_{f,i}$, thereby eliminating the unknown force $R_i$. This results in the following expression for $F_i$:

$$F_i = d_i^2 w\gamma g N_{\gamma,i} + Cwd_iN_{c,i} + C_awd_iN_{a,i} + W_{load,i}N_{q,i}, \tag{1}$$

where $d_i$ is the penetration depth and is found by the distance of point $B$ from stockpile like $OC$. $N_{\gamma,i}$, $N_{c,i}$, $N_{a,i}$, and $N_{q,i}$, are the four bearing capacity factors given as:

$$N_{\gamma,i} = \frac{(\cot\beta_i - \tan\alpha)(\cos\alpha + \sin\alpha\cot(\beta_i+\phi))}{2[\cos(\rho_i+\delta) + \sin(\rho_i+\delta)\cot(\beta_i+\phi)]}, \tag{2}$$

$$N_{c,i} = \frac{1 + \cot\beta_i\cot(\beta_i+\phi)}{\cos(\rho_i+\delta) + \sin(\rho_i+\delta)\cot(\beta_i+\phi)}, \tag{3}$$

$$N_{a,i} = \frac{1 - \cot\rho_i\cot(\beta_i+\phi)}{\cos(\rho_i+\delta) + \sin(\rho_i+\delta)\cot(\beta_i+\phi)}, \tag{4}$$

$$N_{q,i} = \frac{\cos\alpha + \sin\alpha\cot(\beta_i+\phi)}{\cos(\rho_i+\delta) + \sin(\rho_i+\delta)\cot(\beta_i+\phi)}. \tag{5}$$

These dimensionless coefficients account for the influence of different soil properties on the cutting forces. $N_{\gamma,i}$, known as the *unit weight factor*, represents the contribution of the soil's weight to the total force. The *cohesion factor*, $N_{c,i}$, accounts for the influence of soil cohesion, and the *adhesion factor*, $N_{a,i}$ quantifies the contribution of the adhesive forces between the soil and the bucket. Lastly, the *surcharge factor*, $N_{q,i}$, represents the impact of overburden pressure or external loads applied to the soil during excavation. The angle $\beta_i$ is determined by minimizing $N_{\gamma,i}$.

The FEE in its current form suffers from discontinuities under certain geometric configurations. The denominators of the bearing capacity factors in equations (2)–(5) contain cotangent terms, which can become unbounded and introduce numerical instabilities. To overcome this limitation, a set of trigonometric manipulations is introduced to reformulate the bearing capacity factors into a canonical sine-cosine form. This reformulation enables explicit identification of the conditions under which discontinuities occur and provides a means to avoid them. Bounded constraints are imposed during the minimization stage of $N_\gamma$ to ensure a stable and physically valid solution for $\beta$.

To begin this process, the expression for $N_\gamma$ is considered, which can be written as:

$$\cos\alpha + \sin\alpha\cot(\beta_i+\phi) = \frac{\sin(\alpha+\beta_i+\phi)}{\sin(\beta_i+\phi)}, \tag{6}$$

$$\cos(\rho_i+\delta) + \sin(\rho_i+\delta)\cot(\beta_i+\phi) = \frac{\sin(\rho_i+\delta+\beta_i+\phi)}{\sin(\beta_i+\phi)}. \tag{7}$$

By crossing out $\sin(\beta_i + \varphi)$ and recognizing that it must remain nonzero, $N_\gamma$ can be rewritten as:

$$N_\gamma = \frac{(\cot\beta_i - \tan\alpha)\sin(\alpha+\beta_i+\phi)}{2\sin(\rho_i+\delta+\beta_i+\phi)}. \tag{8}$$

Additionally, using

$$\cot\beta_i - \tan\alpha = \frac{\cos(\alpha+\beta_i)}{\cos\alpha\cos\beta_i}, \tag{9}$$

$N_\gamma$ will become:

$$N_{\gamma,i} = \frac{\cos(\alpha+\beta_i)\sin(\alpha+\beta_i+\phi)}{2\cos\alpha\sin\beta_i\sin(\rho_i+\delta+\beta_i+\phi)}. \tag{10}$$

Similarly, $N_c$, $N_a$, and $N_q$ can be rewritten as:

$$N_{c,i} = \frac{\cos\phi}{\sin\beta_i\sin(\rho+\delta+\beta_i+\phi)}, \tag{11}$$

$$N_{a,i} = -\frac{cos(\rho+\beta_i+\phi)\sin(\beta_i+\phi)}{\sin\rho_i\sin(\beta_i+\phi)\sin(\rho_i+\delta+\beta_i+\phi)}, \tag{12}$$

$$N_{q,i} = \frac{\sin(\alpha+\beta_i+\phi)}{\sin(\rho_i+\delta+\beta_i+\phi)}. \tag{13}$$

To ensure continuous force outputs in the FEE-based model, certain trigonometric terms in the denominators of equations (10)–(13) must remain nonzero. Therefore, the following condition must hold: $\sin(\beta_i) \neq 0$, $\sin(\rho_i) \neq 0$, $\cos(\alpha) \neq 0$, $\sin(\beta_i + \phi) \neq 0$, and $\sin(\rho_i + \delta + \beta_i + \phi) \neq 0$. However, these conditions alone are insufficient in practice. Due to temporal discretization, a zero-crossing may occur between time steps without being explicitly captured, which can lead to discontinuities or numerical instabilities in the computed forces. To mitigate this possibility, each of these trigonometric terms is enforced to satisfy $|\cdot| > \varepsilon$, where $\varepsilon$ is a positive threshold.

Some of the above conditions are inherently satisfied by environmental constraints. The stockpile angle $\alpha$ is limited by the material's angle of repose, typically below 45°, ensuring $\cos\alpha > 0.7$. The blade penetration angle $\rho$ must also stay above a minimum threshold as near-zero values make the blade nearly parallel to the surface, invalidating the soil wedge assumption in the FEE model (see Fig. 2). In these simulations, $\rho$ is constrained to remain above 10°.

The remaining conditions involve the terms $\sin(\beta_i)$, $\sin(\beta_i+\phi)$, and $\sin(\rho_i+\delta+\beta_i+\phi)$. As mentioned, bounded constraints are imposed on $\beta_i$ during the minimization of $N_\gamma$. A valid soil wedge requires $\beta_i$ to be strictly positive. Given typical internal friction angles $\phi$ between 22° and 40° (see Table A1), the term $\sin(\beta_i + \phi)$ remains well behaved. However, $\sin(\rho_i + \delta + \beta_i + \phi)$ can approach zero near $\pi$, leading to numerical instability. The set of constraints to implement in the minimization of $N_\gamma$ will become:

$$\beta_i > \varepsilon_1, \quad |\rho_i + \delta + \beta_i + \phi - \pi| > \varepsilon_2, \tag{14}$$

where $\varepsilon_1$ and $\varepsilon_2$ are small positive margins, both set as constant values of 5°. The choice of these $\varepsilon$ values is based on the frequency of data collection and changes in the angles in (14).

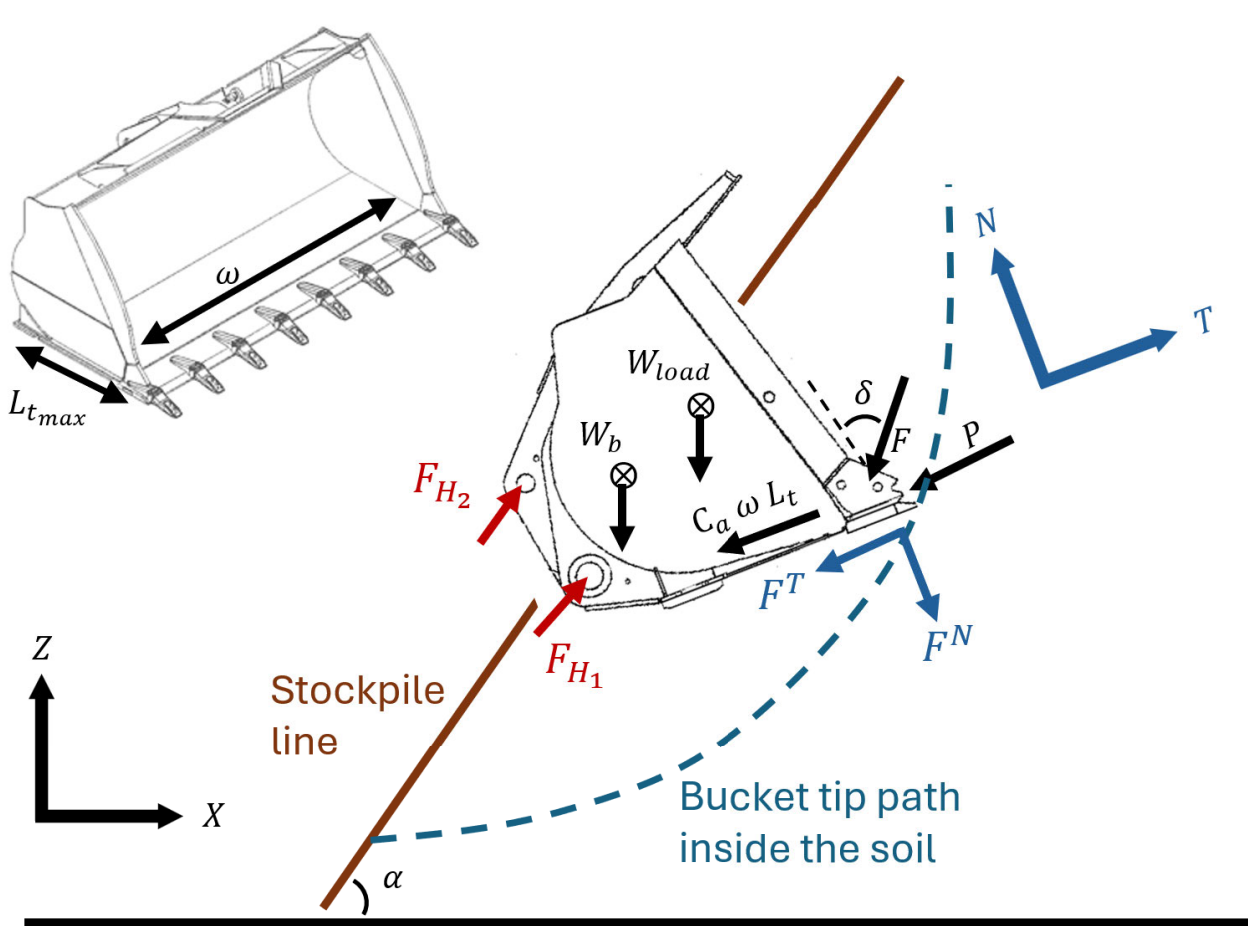

**FIGURE 3. The bucket and all the forces acting on it during the bucket-loading phase.**

Note that there are additional assumptions associated with using the FEE model. As with most quasi-static soil models, the FEE-based force resistance model assumes a slow and steady bucket motion. Although this assumption simplifies real-time implementation, it does not account for dynamic soil displacement effects. Incorporating dynamic geometry updates during excavation remains an important direction for future enhancement of the model.

### B. FORCES ON THE BUCKET DURING BUCKET LOADING

The next step is to model the bucket blade as the cutting blade shown in Fig. 2 with a soil wedge forming along and inside the bucket. The forces acting on the bucket are analyzed, including the resultant force derived from the FEE and the normal compaction force modeled using Bekker's load–sinkage formulation [31].

Inspired by [29] and [30], all forces acting on the bucket blade during the bucket-loading process are illustrated in Fig. 3. The total force includes the reaction force $F_i$, referred to in this work as the FEE force, the adhesion force between the soil and the blade $C_a w L_{t,i}$, and the penetration pressure $P_i$. The penetration pressure $P_i$ is given by:

$$P_i = (\frac{k_c}{b} + k_\phi) d_i^n, \tag{15}$$

where $b$ is the thickness of the cutting edge, and $k_c$, $k_\phi$ are the cohesive and frictional modulus of deformation and $n$ is the exponent of the soil deformation. The resulting forces on the bucket blade can be written in normal tangential coordinates, $F_i^T$ and $F_i^N$ as follows:

$$F_i^T = wbP_i + F_i \sin\delta + C_a w L_{t,i}, \tag{16}$$

$$F_i^N = F_i \cos\delta. \tag{17}$$

These forces are related to the forces $F_{H_1}$ and $F_{H_2}$ measured through the load cells on the hinges as shown in Fig. 3. The weight of the bucket $W_b$ is known since the weight of the soil loaded into the bucket is estimated using its positioning and the geometry of the stockpile.

## III. PARAMETER OPTIMIZATION METHOD

Following the establishment of the FEE model for bucket-soil interaction in Section II, the optimization of soil parameter values is performed. As listed in Table 1, there are eight soil parameters, $\theta$, that need to be optimized to model excavation forces:

$$\theta = \left[\gamma \;\; C \;\; C_a \;\; \phi \;\; \delta \;\; k_c \;\; k_\phi \;\; n\right]^T. \tag{18}$$

The overall parameter optimization framework operates by fitting the analytical model described in Section II to force data measured on the bucket. The first required input is the bucket's position and orientation obtained either from onboard sensors (e.g., an inertial measurement unit (IMU) mounted on the bucket) or inferred through cylinder strokes using a kinematic model of the front-end mechanism [34]. This geometric information defines the soil wedge configuration at each time step. The second input consists of the hinge forces, $F_{H_1}$ and $F_{H_2}$, shown in Fig. 3. These measurements are resolved into the normal and tangential components denoted as $F_{\text{obs}}^N$ and $F_{\text{obs}}^T$ and serve as the observed forces. The framework then applies an optimization-based approach to estimate the soil parameters $\theta$ by minimizing the discrepancy between observed and analytically predicted bucket forces. Unlike linear systems, where least squares methods are sufficient, the nonlinear soil-tool interaction model requires nonlinear optimization. This optimization arises from nested trigonometric expressions, force-angle dependencies, and geometry-driven terms as well as implicit and constrained relationships between soil parameters and output forces.

The parameter optimization problem is formulated by treating the eight soil-related quantities as independent variables. However, the high dimensionality and nonlinearity of the force model make simultaneous optimization prone to poor convergence and inefficiency. To overcome this limitation, a multi-stage optimization approach is introduced. Mathematical analysis of the model equations reveals a natural separation of terms, allowing the parameters to be grouped into distinct subsets and optimized sequentially.

To evaluate the effectiveness of the proposed method, and given the lack of comparable data-efficient approaches, a baseline strategy is introduced. This baseline consists of a single-stage optimization in which all eight soil parameters are estimated simultaneously, serving as a reference for assessing the multi-stage formulation. The problem is posed as a constrained, nonlinear least squares optimization and solved using the gradient-based algorithm L-BFGS-B [35], which accommodates bounded constraints. Since the model is non-convex, the optimization is sensitive to initial parameter values and may converge to local minima. To mitigate this issue, several strategies were tested, including warm-starting, randomized initializations, and multiple starting points. Warm-starting was examined in two forms: initialization from values within the soil parameter ranges and reuse of estimated parameters from the previous cycle. These approaches improved convergence. Future work will focus on a systematic sensitivity analysis of soil parameters and initialization choices.

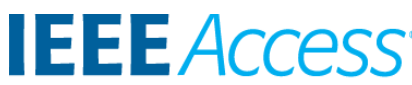

### A. SINGLE-STAGE FULL MODEL OPTIMIZATION

For the baseline single-stage optimization, the optimization problem is formulated as follows:

$$\min_{\theta} \quad J_\theta = \lambda \sum_{i=1}^{N} (F^T_{\text{obs},i} - F^T_i)^2 + (1-\lambda) \sum_{i=1}^{N} (F^N_{\text{obs},i} - F^N_i)^2 \tag{19a}$$

$$\text{s.t.} \quad F^T_i = wbP_i + F_i \sin\delta + C_a w L_{t,i}, \tag{19b}$$

$$F^N_i = F_i \cos\delta, \tag{19c}$$

$$P_i = \left(\frac{k_c}{b} + k_\phi\right) d_i^n, \tag{19d}$$

$$F_i = d_i^2 w\gamma g N_{\gamma,i} + C w d_i N_{c,i} + C_a w d_i N_{a,i} + W_{\text{load}} N_{q,i}, \tag{19e}$$

$$N_{\gamma,i} = f_\gamma(\alpha, \beta_i, \phi, \rho_i, \delta), \tag{19f}$$

$$N_{c,i} = f_c(\beta_i, \phi, \rho_i, \delta), \tag{19g}$$

$$N_{a,i} = f_a(\beta_i, \phi, \rho_i, \delta), \tag{19h}$$

$$N_{q,i} = f_q(\alpha, \beta_i, \phi, \rho_i, \delta), \quad i = 1, \ldots, N \tag{19i}$$

$$\theta_{\min} \leq \theta \leq \theta_{\max}. \tag{19j}$$

Here, $F^T_i$ and $F^N_i$ represent the predicted tangential and normal forces from the model, while $F^T_{\text{obs},i}$ and $F^N_{\text{obs},i}$ denote the corresponding observed (measured) values obtained from sensor data. The objective function $J_\theta$ is defined as a weighted sum of two mean squared error terms, corresponding to the discrepancies in these horizontal and normal force predictions. The weighting parameter $\lambda$ is introduced to control the relative contribution of each term. A $\lambda = 0.5$ is assumed to keep the contribution of terms balanced.

The constraint (19b) defines the horizontal force based on equation (16), while constraint (19c) specifies the normal force following equation (17). The total resistive force from FEE is incorporated through constraint (19e) as defined in equation (1). The bearing capacity factors, derived in equations (2)–(5), appear explicitly in (19f)–(19i). To ensure physically meaningful results, the soil parameters are bounded using hard constraints based on ranges reported in the literature and summarized in Table A1 and Table A2. For density $\gamma$, the range was set to 1297–2345 kg/m$^3$. Cohesion $C$ and adhesion $C_a$ were both constrained to 0–50000 N/m$^2$, which encompasses soft to firm and non-cohesive to cohesive soil types. The internal and external friction angles were bounded between 0° and 45°, corresponding to 0.0 to 0.785 rad. For Bekker's pressure–sinkage parameters, the allowed ranges were $k_c = 0.00$ to 10.0 kN/m$^{n+1}$, $k_\phi = 0$ to 5000 kN/m$^{n+2}$, and exponent $n = 0.11$ to 1.53. These limits are consistent among all the parameter estimation approaches done in this paper.

### B. MULTI-STAGE OPTIMIZATION

When optimizing a large number of parameters in a complex non-linear model, solving for all parameters simultaneously can result in poor convergence and high computational cost. To overcome this possibility, a sequential optimization approach is adopted that decomposes the parameter estimation problem into multiple stages. This decomposition is enabled by analyzing the form of the force balance equations in (15), (16), and (17), which reveal separable dependencies among parameter groups. The process begins with the normal component of the bucket force, $F^N_{obs}$, obtained directly from measurements, and uses it to reconstruct the FEE force $F$ using equation (17):

$$F^N_{obs} = F\cos\delta, \qquad F_{obs} = \frac{F^N_{obs}}{\cos\delta}. \tag{20}$$

Here, $F_{obs}$ represents the FEE-equivalent total force reconstructed from the observed normal force. Substituting equations (20) and (15) into the tangential force expression (16) yields:

$$F^T = wb\left(\frac{k_c}{b} + k_\phi\right) d^n + F^N_{obs}\tan\delta + C_a w L_t. \tag{21}$$

From this formulation, a subset of parameters $\theta_1$ is identified:

$$\theta_1 = \left[C_a \ \delta \ k_c \ k_\phi \ n\right]^T. \tag{22}$$

The force $F^T$ in equation (21), constructed from the observed normal force $F^N_{\text{obs}}$, forms the basis of Stage 1 optimization. This stage aims to minimize the difference between the predicted tangential force from (21) and the observed value. The cost function $J^1_{\theta_1}$ and its associated constraints are defined as:

$$\min_{\theta_1} \quad J_{\theta_1} = \sum_{i=1}^{N} (F^T_{obs,i} - F^T_i)^2 \tag{23a}$$

$$\text{s.t.} \quad F^T_i = wb\left(\frac{k_c}{b} + k_\phi\right) d_i^n + F^N_{obs,i}\tan\delta + C_a w L_{t,i}, \quad i = 1, \ldots, N, \tag{23b}$$

$$\theta_{1,\min} \leq \theta_1 \leq \theta_{1,\max}. \tag{23c}$$

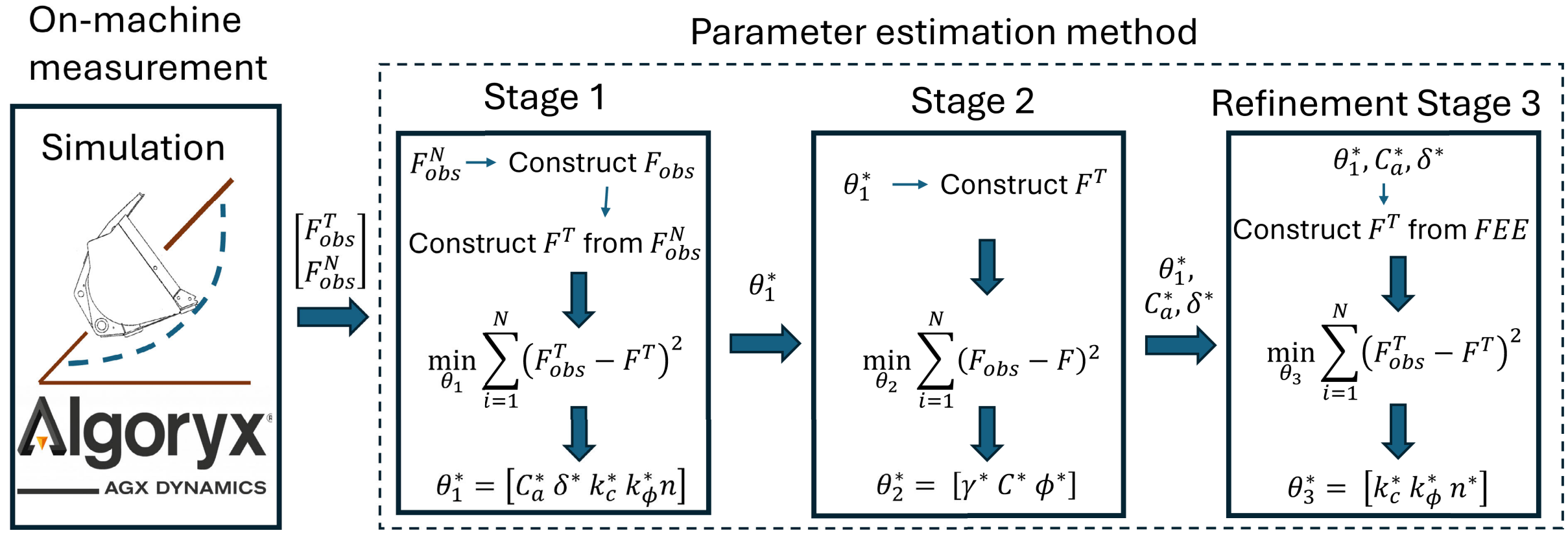

**FIGURE 4.** **Overview of the proposed multi-stage parameter estimation framework. Force data ($F_{\text{obs}}^T$, $F_{\text{obs}}^N$) from a real bucket-loading cycle is used to estimate parameters in three sequential stages. In Stage 1, $F_{\text{obs}}^N$ is used to reconstruct the FEE force $F$, which is then used to model $F^T$ and identify first subset of parameters $\theta_1$. In Stage 2, $\theta_1^*$ is used to model the full FEE force $F$, enabling estimation of material parameters $\theta_2$. In Stage 3, a refinement step is introduced to re-optimize the subset $\theta_3$ that influences $F^T$, improving tangential force accuracy while keeping other parameters fixed.**

By solving this optimization, the first set of optimized parameters $\theta_1^*$ is obtained:

$$\theta_1^* = \left[C_a^* \; \delta^* \; k_c^* \; k_\phi^* \; n^*\right]^T. \tag{24}$$

Stage 2 builds on the previously obtained solution $\theta_1^*$ by fixing its values and incorporating them into a second stage formulation. Using the optimal parameter $\delta^*$ and (20), the reconstructed FEE force $F_{\text{obs}}$ can be computed based on the normal force $F_{\text{obs}}^N$:

$$F_{obs} = \frac{F_{obs}^N}{\cos \delta^*}. \tag{25}$$

Eq. (25) and (1) form the foundation for the second stage of parameter optimization. This stage involves optimizing the difference between the FEE force estimated through the empirical approach in (1) and the observed data, using the subset of parameters determined in the first stage as in (25). The cost function is formulated as a non-linear least squares objective minimizing the difference between observed and predicted FEE forces. The second subset of parameters is:

$$\theta_2 = \left[\gamma \; C \; \phi\right]^T. \tag{26}$$

The full set of cost function and constraints is given below:

$$\min_{\theta_2} \; J_{\theta_2} = \sum_{i=1}^{N} \left(F_{\text{obs},i} - F_i\right)^2 \tag{27a}$$

$$\text{s.t.} \; F_i = d_i^2 w \gamma g N_{\gamma,i} + C w d_i N_{c,i} + C_a^* w d_i N_{a,i} + W_{\text{load}} N_{q,i}, \tag{27b}$$

$$N_{\gamma,i} = f_\gamma(\alpha, \beta_i, \phi, \rho_i, \delta^*), \tag{27c}$$

$$N_{c,i} = f_c(\beta_i, \phi, \rho_i, \delta^*), \tag{27d}$$

$$N_{a,i} = f_a(\beta_i, \phi, \rho_i, \delta^*), \tag{27e}$$

$$N_{q,i} = f_q(\alpha, \beta_i, \phi, \rho_i, \delta^*), \quad i = 1, \ldots, N, \tag{27f}$$

$$\theta_{2,\min} \leq \theta_2 \leq \theta_{2,\max}. \tag{27g}$$

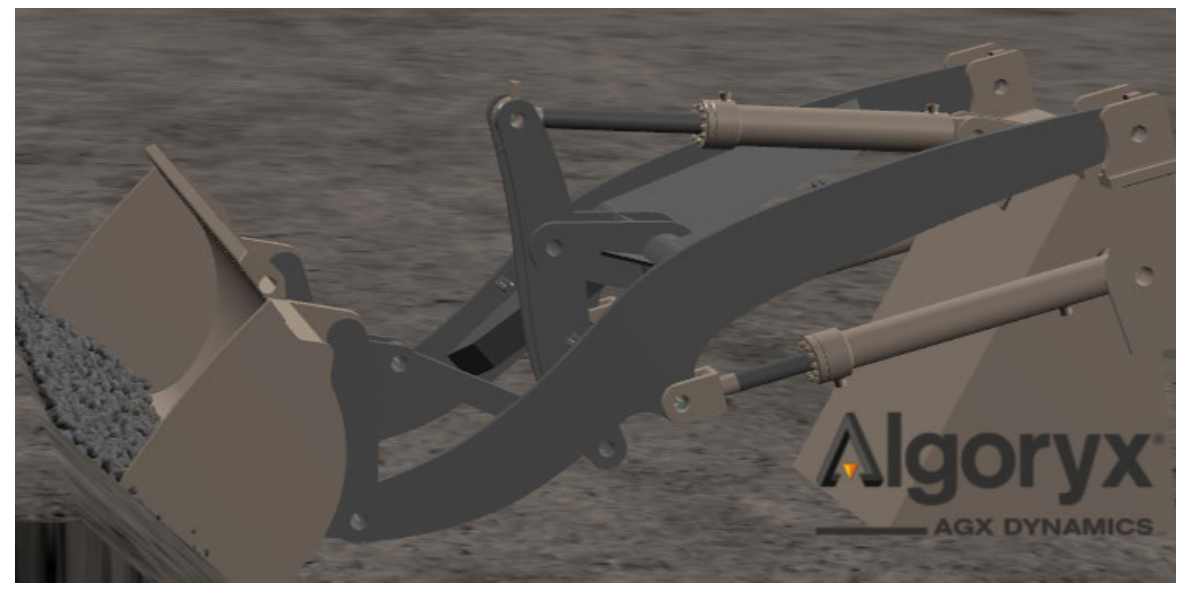

**FIGURE 5.** **Digital twin model of the EVERUN ER12 wheel loader's front-end assembly developed in Algoryx Dynamics [36].**

In this optimization formulation, the cost function minimizes the difference between observed force (25) and $F_i$, which is defined in (27b). Unlike the baseline optimization, (27b) is defined with the fixed parameter $C_a^*$. The four bearing capacity factors, given in (27c)–(27f), are computed following the same methodology as in (2)–(5). Similarly, the external friction angle $\delta^*$ is treated as a fixed parameter derived from the optimization results in the preceding stage.

The multi-stage optimization results revealed that the root mean square error (RMSE) for the normal force $F^N$ was consistently lower than that for the tangential force $F^T$. This discrepancy aligns with the initial assumption whereby the total force $F$ was reconstructed from the observed normal force $F_{\text{obs}}^N$, inherently biasing the estimation process toward improved accuracy in the normal direction. To address this imbalance and further reduce error in the tangential force, a third refinement step is introduced. In this stage, a reduced-order modeling approach is used to re-optimize the subset of parameters that predominantly influence the tangential component, while the remaining parameters are fixed at the values obtained in the earlier stages. The subset of parameters was identified by analyzing the governing equations and isolating those that influence the tangential force $F^T$ but not the normal force $F^N$. From equations (16) and (17), the compaction pressure term $P$, modeled using Bekker's load-sinkage formulation, is shown to exclusively affect $F^T$. The relevant parameters are $k_c$, $k_\phi$, and $n$. Optimization in this stage follows the baseline formulation with $\lambda = 1$, while all remaining parameters are held constant at their values from earlier stages. The final subset of parameters re-optimized in this refinement step is:

$$\theta_3 = \begin{bmatrix} k_c \ k_\phi \ n \end{bmatrix}^T . \tag{28}$$

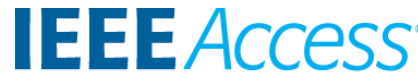

**TABLE 2.** Bulk and particle properties for gravel, sand, dirt extracted from Algoryx Terrain and modified sand. The modified sand is manually changed to magnify the change in density and friction angle to test the proposed methods sensitivity to different values.

| Property | Gravel | Sand | Dirt | Modified Sand |
|---|---|---|---|---|
| **Bulk Properties** | | | | |
| Cohesion (Pa) | 0.0 | 0.0 | 2100 | 0.0 |
| Density (kg/m$^3$) | 1474.0 | 1474.0 | 1474.0 | 1874.5 |
| Dilatancy Angle (radians) | 0.19188 | 0.157 | 0.2267 | 0.157 |
| Friction Angle (radians) | 0.7679 | 0.6807 | 0.6978 | 0.5934 |
| Maximum Density (kg/m$^3$) | 1600.0 | 1800.0 | 2000.0 | 2100.0 |
| Poisson's Ratio | 0.1 | 0.1 | 0.15 | 0.1 |
| Swell Factor | 1.0 | 1.0 | 1.1 | 1.0 |
| Young's Modulus (Pa) | 4.6E+6 | 4.5E+6 | 6.5E+6 | 4.6E+6 |
| Delta Repose Angle (radians) | 0.0 | 0.0 | 0.0 | 0.0 |
| **Particle Properties** | | | | |
| Particle Cohesion (Pa) | 0.0 | 0.0 | 2500.0 | 0.0 |
| Particle Rolling Resistance | 0.3 | 0.1 | 0.1 | 0.1 |
| Particle-Terrain Cohesion (Pa) | 0.0 | 0.0 | 2500.0 | 0.0 |
| Particle-Terrain Friction | 0.9 | 0.7 | 0.7 | 0.7 |
| Particle-Terrain Restitution | 0.1 | 0.0 | 0.0 | 0.0 |
| Particle-Terrain Rolling Resistance | 0.9 | 0.7 | 0.7 | 0.7 |
| Particle-Terrain Young's Modulus (Pa) | 1E+8 | 1E+8 | 1E+8 | 1E+8 |
| Particle Young's Modulus (Pa) | 1E+7 | 1E+7 | 1E+7 | 1E+7 |

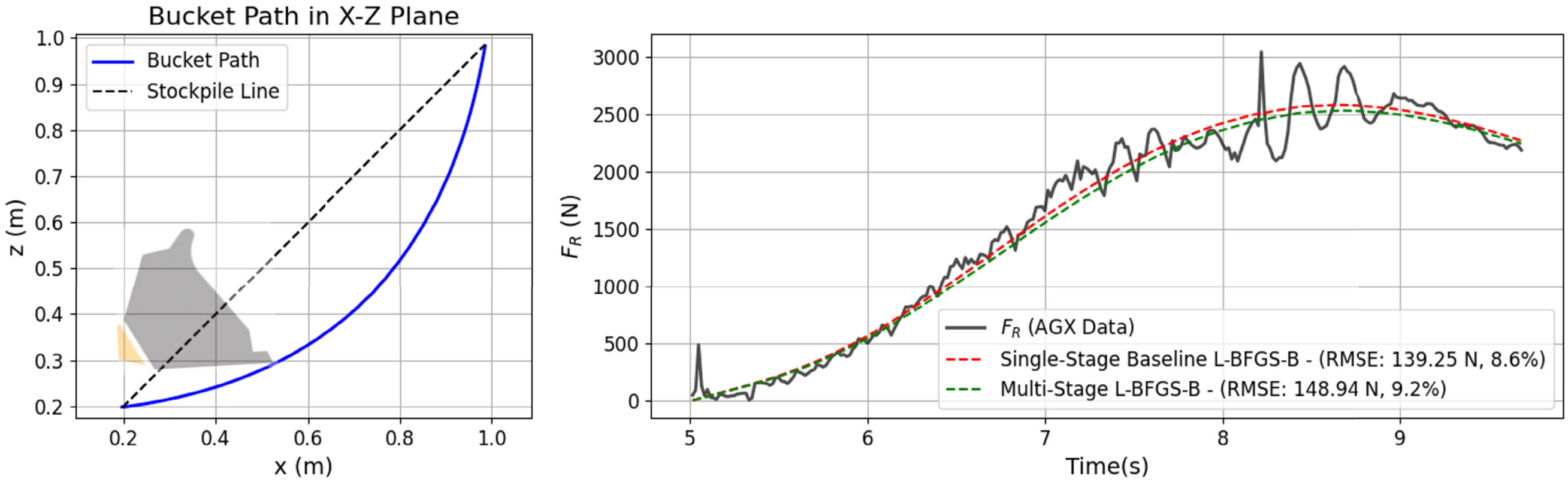

**FIGURE 6.** Comparison of force prediction accuracy between the single-stage and multi-stage parameter optimization methods.(Left) The bucket path in the $x$–$z$ plane during a fixed excavation cycle, with the stockpile line shown for reference.(Right) The resultant force $F_R$ over time is compared against the ground truth from Algoryx Dynamics.

This refinement rebalances the force predictions and improves consistency between the estimated and observed tangential and normal forces.

### C. SIMULATION FRAMEWORK

This study models excavation scenarios using high-fidelity simulations in Algoryx Dynamics, a physics engine developed by Algoryx [37]. A snapshot of the simulation environment in Algoryx Dynamics is shown in Fig. 5, illustrating the wheel loader interacting with the soil pile. Algoryx is tailored to multibody systems with contact and friction dynamics and is suited for simulating articulated machinery in deformable environments. It incorporates non-smooth, multibody dynamics [38], DEM [39], and Hertz-Mindlin contact theory [40] to capture soil–tool interactions.

Soil is simulated using the Algoryx terrain module, which represents soil as collections of spherical particles with six degrees of freedom. These particles interact via Hertzian elasticity, Coulomb friction, and rolling resistance. Material properties are assigned by category, grouped according to their influence on bulk behavior, compaction, particle interactions, and terrain contact. The material properties in the Algoryx terrain are grouped into four categories: *bulk properties*, which govern general behavior such as soil density, failure zone shape, and excavation forces; *compaction properties*, which define the soil's response to compression; *particle properties*, which regulate contact dynamics between soil particles and with the terrain; and *excavation contact properties*, which manage interactions between the shovel, terrain, and soil aggregates. Algoryx Terrain also provides a built-in material library with preset parameters for three common soil types: gravel, sand, and dirt. The bulk and particle properties associated with these presets have been extracted and are detailed in Table 2.

**TABLE 3.** Comparison of parameter optimization methods: single-stage and multi-stage L-BFGS-B. Grouped soil parameters and RMSE values are shown across different optimization stages.

| Method | | Soil Parameters | | | | | | | | Results | | | |
|---|---|---|---|---|---|---|---|---|---|---|---|---|---|
| | | $\gamma$ (kg/m$^3$) | $C$ (N/m$^2$) | $C_a$ (N/m$^2$) | $\phi$ (°) | $\delta$ (°) | $k_c$ (N/m$^{n+1}$) | $k_\phi$ (N/m$^{n+2}$) | $n$ (–) | $F^T$ RMSE (N,%) | $F^N$ RMSE (N,%) | $F_R$ RMSE (N,%) | Time (s) |
| Single-Stage | – | 1600.1 | 152.4 | 0.0 | 45.0 | 36.7 | 748.0 | 163.8 | 1.06 | 96.2(9.0%) | 137.7(11.4%) | 139.2(8.6%) | 87.21 |
| Multi-Stage | – | 1635.1 | 183.5 | 0.0 | 45.0 | 35.1 | 745.6 | 166.9 | 0.91 | 97.3(9.1%) | 139.6(11.6%) | 141.1(8.7%) | 47.00 |
| Multi-Stage | | | | | | | | | | Stage 1 RMSE | Stage 2 RMSE | Stage 3 RMSE | Time (s) |
| | Stage 1 | – | – | 0.0 | – | 35.1 | 993.6 | 558.6 | 0.46 | 103.1(9.6%) | - | – | 20.37 |
| | Stage 2 | 1635.1 | 183.5 | – | 45.0 | – | – | – | – | – | 108.5(7.4%) | - | 16.49 |
| | Stage 3 | – | – | – | – | – | 745.6 | 166.9 | 0.91 | 97.3(9.1%) | - | - | 10.14 |

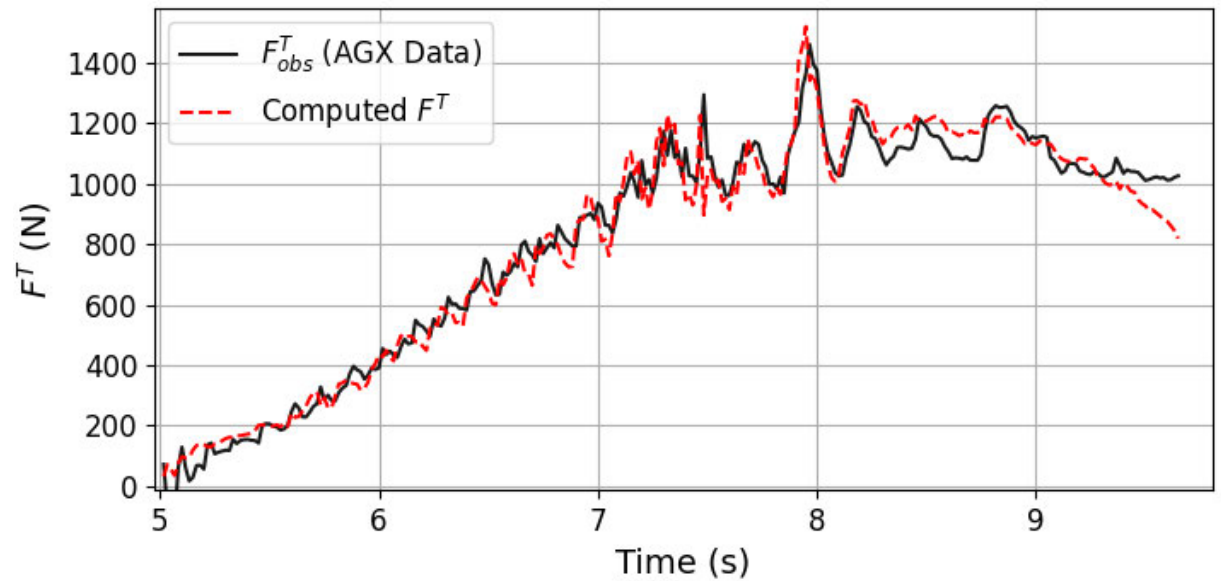

(a)

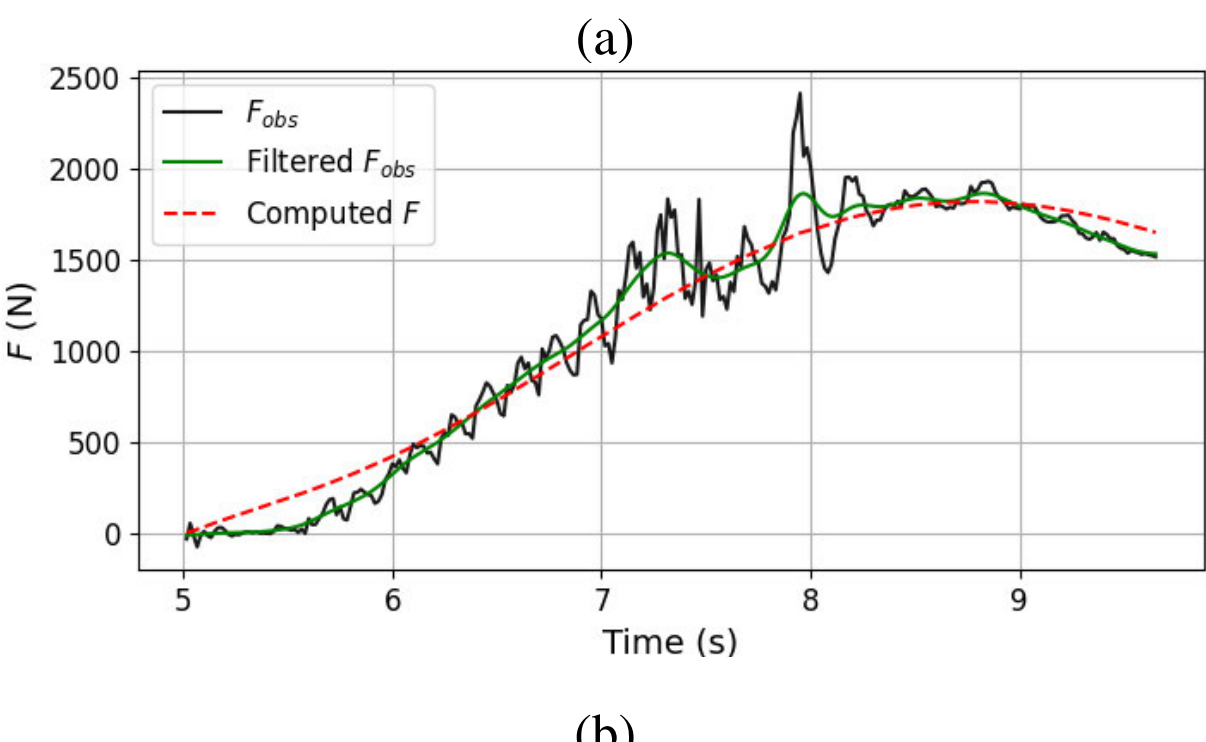

(b)

**FIGURE 7.** Optimization results of the multi-stage parameter estimation method. (a) Stage 1: Comparison of observed and computed horizontal force $F^T$. (b) Stage 2: Comparison of observed, filtered, and computed FEE force $F$. Gaussian filter was applied to the observed force $F_{\text{obs}}$ to reduce the effect of noise and vibration for improved convergence.

Algoryx Dynamics was validated using a calibrated digital twin of an EVERUN ER12 wheel loader developed at the Advanced Highway Maintenance and Construction Technology (AHMCT) Research Center, UC Davis. The physical machine was instrumented with hydraulic pressure sensors, inclinometers, and custom dual-axis load pins at the bucket hinge. Data from real-world operation were used to calibrate and validate the simulation parameters in a prior work [36].

It is important to note that the Algoryx Dynamics terrain model is fundamentally different from this analytical framework. While Algoryx employs a numerical, particle-based approach to simulate soil behavior, this model is analytical and equation-based. As a result, the parameter sets used in each approach differ in both form and purpose. However, a few physical parameters are shared across both models, specifically soil density $\gamma$, cohesion $C$, adhesion $C_a$, and internal friction angle $\phi$.

## IV. RESULTS AND DISCUSSION

The results of this study are structured to evaluate the performance and generalizability of the proposed force estimation framework. The analysis begins with a fixed bucket loading cycle, defined by a consistent excavation trajectory in the $x-z$ plane, with a given bucket path, orientation, and speed profile.

The multi-stage optimization approach is first compared with the baseline method introduced in Section III-A, with both solved using the L-BFGS-B algorithm [35]. For consistency, all optimization runs used the same hyper-parameters: 1000 maximum iterations, a gradient norm tolerance of $10^{-5}$, bounded parameter ranges, and gradient approximations via finite differences. All methods were initialized from the same guess and executed on the same machine (14th-gen Intel Core i9). To assess generalizability, a train–test strategy is employed. Parameters were optimized in one excavation cycle (''train'' phase) and then used to predict forces on separate, unseen paths (''test'' phase). In addition, a dual-cycle scenario is evaluated whereby parameters estimated from an initial excavation are applied to a subsequent cycle. To examine sensitivity to terrain, the parameter optimization process is repeated under different soil types while keeping the trajectory fixed. Model performance is evaluated using the RMSE of the resultant force, defined as $F_R = \sqrt{(F^T)^2 + (F^N)^2}$, compared to the ground-truth resultant force $F_{R,\text{obs}}$ obtained from Algoryx simulation.

### A. SINGLE PATH MODEL COMPARISON

To evaluate the multi-stage parameter optimization method against the baseline approach, a single bucket loading scenario is simulated. This scenario captures a full excavation cycle from soil entry to bucket exit and records bucket force data at a frequency of 60 Hz over 4.67 seconds, resulting in 281 data points. The soil is modeled as a trapezoidal stockpile composed of gravel, with material properties listed in Table 2. The trajectory of the bucket tip during this cycle is shown in the left panel of Fig. 6. This ''single path'' represents a typical smooth bucket-loading motion.

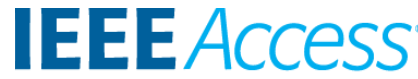

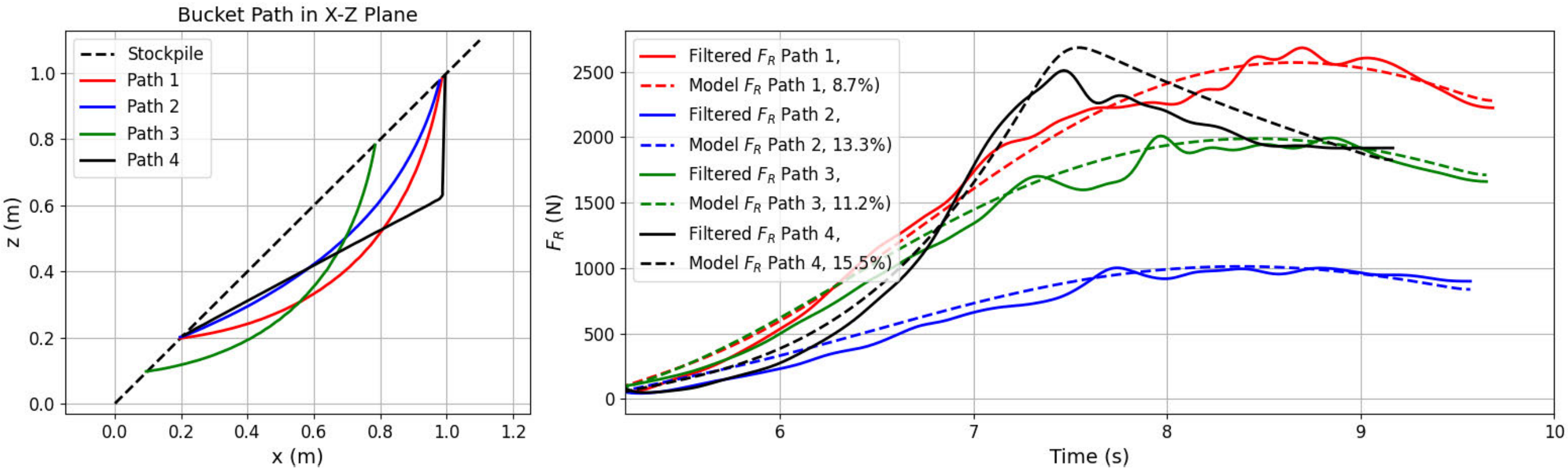

**FIGURE 8.** Comparison of bucket tip trajectories (left) and resultant forces $F_R$ (right) for four excavation paths. The dashed black line in the left plot represents the stockpile slope, while each colored path shows a different digging trajectory. On the right, solid lines represent filtered (observed) forces, and dashed lines show the model's predicted forces, with the corresponding RMSE values listed in the legend.

**TABLE 4.** Soil parameters from Algoryx (AGX) and parameter estimation (PE) for different soil types using the Multi-stage optimization approach.

| Soil Parameters | Gravel | | Modified Sand | | Dirt | |
|---|---|---|---|---|---|---|
| | AGX | PE | AGX | PE | AGX | PE |
| $\gamma$ (kg/m$^3$) | 1474.0 | 1513.4 | 1874.5 | 1802.3 | 1474.0 | 1833.17 |
| $C$ (N/m$^2$) | 0.0 | 143.8 | 0.0 | 216.0 | 2100.0 | 0.00 |
| $C_a$ (N/m$^2$) | 0.0 | 184.0 | 0.0 | 72.4 | 2500.0 | 3039.52 |
| $\phi$ (rad) | 0.76 | 0.78 | 0.59 | 0.78 | 0.69 | 0.37 |
| $\delta$ (rad) | N/A | 0.63 | N/A | 0.66 | N/A | 0.39 |
| $k_c$ (N/m$^{n+1}$) | N/A | 1274.0 | N/A | 1807.6 | N/A | 5486.1 |
| $k_\phi$ (N/m$^{n+2}$) | N/A | 177.4 | N/A | 188.0 | N/A | 261.3 |
| $n$ (-) | N/A | 1.21 | N/A | 1.50 | N/A | 1.44 |
| $F_R$ RMSE N(%) | N/A | 188.2 (9.0%) | N/A | 224.1 (8.6%) | N/A | 540.6 (15.3%) |
| time (s) | N/A | 44.05 | N/A | 38.23 | N/A | 41.37 |

The results of parameter optimization using single-stage and multi-stage optimization are summarized in Table 3. For the single-stage method, the RMSE for tangential and normal forces $F^T$ and $F^N$ were 9.0% and 11.4%, respectively, with a resultant force $F_R$ RMSE of 8.6%. The total convergence time was 87.21 seconds. The multi-stage method produced a slightly higher RMSE of 9.1% for $F^T$, 11.6% for $F^N$, and 8.9% for $F_R$ but reduced the total optimization time by 46%, converging in 47.0 seconds across three stages. This trade-off favors multi-stage optimization in scenarios requiring rapid, cycle-to-cycle adaptation, such as real-time autonomous excavation where computation resources and response times are constrained.

Compared to existing methods, the multi-stage approach offers competitive accuracy with greater data efficiency. For example, Yu et al. [19] reported a vertical force RMSE of 18.92% using a pure FEE model and 10.54% using a hybrid FEE–RNN model trained on 10 real excavation cycles. While their hybrid method improves accuracy, it relies on large datasets and offline training. This framework achieves similar performance using only a single excavation cycle and requires no learning infrastructure, making it more practical for online implementation. Similarly, Luengo et al. [17] reformulated the classical FEE model to account for slope effects and soil remolding forces, reducing force prediction error from approximately 30,000 N to 13,000 N across eight excavation cycles. However, their parameter estimation relied on computationally intensive methods with exhaustive searches requiring up to 20 hours and hybrid simulated annealing–Powell optimization still taking several minutes. In contrast, the proposed framework achieves accurate force prediction from a single cycle gradient-based optimization. Tan et al. [18] also explored online soil parameter estimation using a Newton–Raphson algorithm applied to Mohr–Coulomb and Chen–Liu Upper Bound (CLUB) models, achieving up to 3500 times faster computation compared to the Parameter Space Intersection Method (PSIM). While their approach demonstrated robust convergence, it required repeated excavation trials for stable estimation and relied on models with fewer parameters due to simplified geometric assumptions, limiting their ability to capture complex soil–tool interactions. By comparison, the proposed multi-stage framework maintains analytical tractability, avoids repeated trials or tables lookups, and delivers efficient force prediction from a single excavation cycle.

Scenario characteristics influence model performance. In this case, the smooth trajectory contributes to better accuracy of FEE model predictions. While this may account for the lower RMSE values compared to other studies, it aligns with the objective: favoring energy-optimal, impact-free bucket motions. Since sudden force transitions are undesirable in autonomous excavation, smoother soil pickups are both a modeling preference and an operational goal.

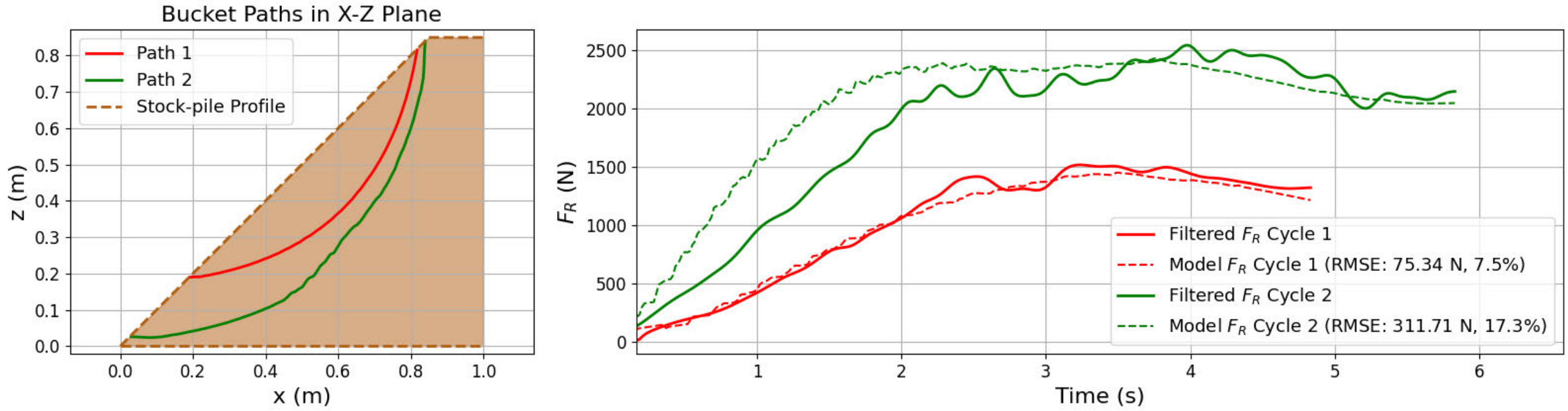

**FIGURE 9. (Left) Bucket tip trajectory during a dual-cycle excavation in a trapezoidal soil pile. The blue line shows the full motion path of the bucket, while red and green segments indicate Cycle 1 and Cycle 2 respectively. (Right) Comparison between filtered resultant force $F_R$ from simulation and the model prediction. Parameters were estimated using data from Cycle 1 and reused for Cycle 2 to evaluate generalization performance.**

The results of each stage of the multi-stage optimization are illustrated in Fig. 7. In Stage 1, a subset of parameters $\theta_1$ is optimized without utilizing the full FEE model; instead, the FEE force $F_i$ is reconstructed directly from the observed normal force $F^N_{\text{obs}}$. As a result, the optimization is less sensitive to measurement noise, and no filtering is required since both the input and output forces are derived from simulation data. This robustness is evident in Fig. 7a where close agreement is observed between predicted and reference forces even in the absence of pre-processing. This stage is also the most computationally intensive as it involves estimating five parameters, which is more than any subsequent stage.

In Stage 2, the optimized parameters $\theta_1^*$ from Stage 1 are held fixed while the remaining parameters are identified by fitting the full FEE model onto simulation data. To enhance convergence and mitigate the effects of noise and vibrations, Gaussian filtering is applied (see Fig. 7b). The RMSE for $F_T$ remains elevated in this scenario, primarily due to the behavior of the compaction term $P$, which strongly influences tangential forces. This continued elevation reflects a known challenge when modeling force components governed by nonlinear sinkage mechanics and is within the limits of the FEE's quasi-static formulation. A refinement stage is, therefore, introduced to improve this fit while keeping the model's structure analytically tractable.

Stage 3 re-optimizes only the compaction-related parameters, while keeping all other parameters fixed at their previously identified values. This targeted refinement is performed using the same optimization framework described for the baseline single-stage method, but with a weighting factor of $\lambda = 1$, as the focus is solely on improving the tangential force prediction. Since only a subset of the full parameter set is optimized, this step constitutes a reduced-order modeling approach. The refinement leads to a substantial improvement in accuracy, reducing the RMSE to 7.6% for $F^T$, 7.0% for $F^N$, and 6.8% for $F_R$, with only an additional 10.14 seconds of computation time.

To qualitatively compare the performance of the proposed parameter optimization methods, the resulting force is shown in the right panel of Fig. 6. Both approaches closely follow the general force trend; although, the single-stage method slightly outperforms the multi-stage method in peak force alignment and overall RMSE.

### B. MULTIPLE PATHS

As the aim of this study was to enable reliable force estimation for an upcoming excavation cycle, the analysis focuses on a single soil profile and applies the multi-stage optimization approach. As shown in Fig. 8, each trajectory represents a unique digging motion over the same stockpile, varying in bucket orientation and depth. Trajectories were generated using a quadratic Bézier formulation, with curvature control parameters adjusted to create (1) varying curvatures, (2) different start/end points, and (3) segmented linear paths. The left panel shows one example of each trajectory in the $x$–$z$ plane, and the right compares predicted and observed resultant forces $F_R$.

In this evaluation, Path 1 is used for parameter estimation, while Paths 2–4 serve as test cases using the identified parameters. The model achieves RMSE values between 11.2% and 15.5% across all paths. Notably, Path 3, despite differing in start and end points, shares the same motion profile as Path 1 and yields the best accuracy. This result suggests that the similarity in motion dynamics influences generalization more than spatial position. As trajectories deviate further from the original, prediction error increases but remains below 15.5%, indicating that trajectory-dependent variation is a key challenge. While the model captures overall force trends, timing mismatches in peak forces reveal limitations in modeling transient dynamics, such as abrupt changes in penetration rate or tool–soil contact. These issues likely arise from the quasi-static FEE formulation, which lacks velocity or acceleration dependence. Incorporating dynamic terms—such as rate-sensitive contact mechanics or inertial effects—could improve the model's responsiveness to rapid transitions and enhance its predictive accuracy during transient excavation phases.

One aspect not captured in the current results is the effect of changes in stockpile geometry caused by previous excavation passes, which may impact the predictive accuracy of the model for subsequent trajectories. To evaluate this effect, a dual-cycle excavation scenario is considered. In the first cycle, the bucket enters and exits the stockpile, and the corresponding force data are used for parameter estimation. This scenario is visualized in Fig. 9 where the soil pile is shown in brown, and the trajectory of the bucket tip is shown in red in the first cycle and green in the second cycle. The results of this run demonstrate that the parameter optimization method achieves an RMSE of 7.5% for Cycle 1 using the identified parameters: soil density $\gamma = 1855\,\text{kg/m}^3$, cohesion $C = 194.25\,\text{N/m}^2$, adhesion $C_a = 0\,\text{N/m}^2$, internal friction angle $\phi = 45°$, external friction angle $\delta = 35.37°$, and compaction parameters $k_c = 1298\,\text{N/m}^{n+1}$, $k_\phi = 250\,\text{N/m}^{n+2}$, and $n = 1.50$.

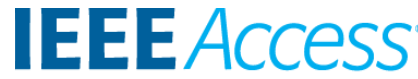

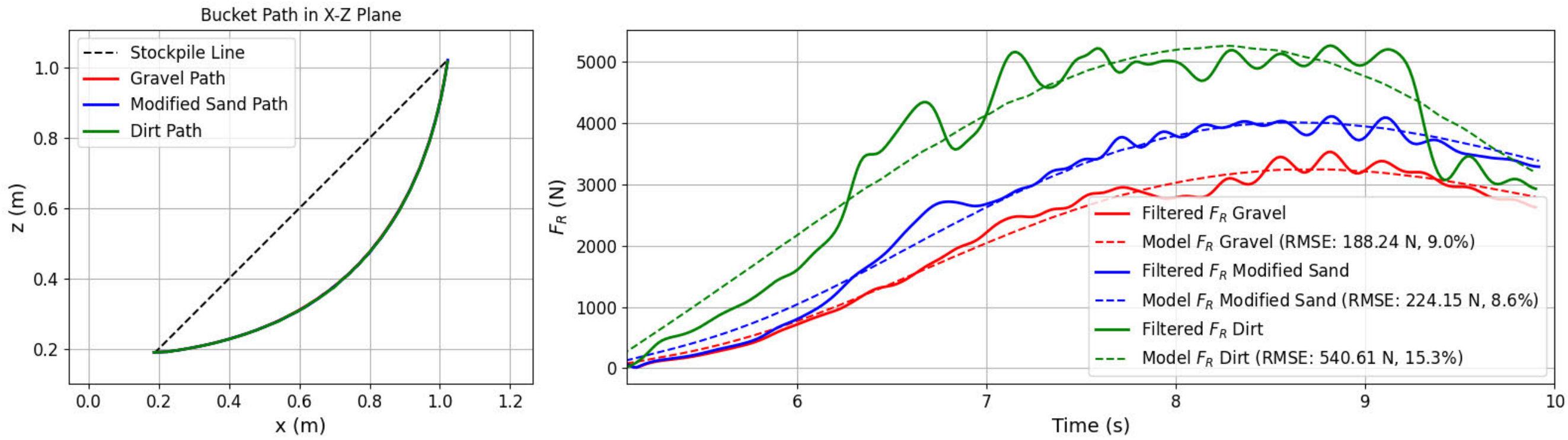

**FIGURE 10. Comparison of resultant forces $F_R$ for a given excavation path for 3 different soil types. The different soils and parameters alongside found parameters are listed in Table 4.**

The identified parameters are then used to predict forces for the second cycle as shown in the right panel of Fig. 9. However, after the first excavation cycle, the soil surface is reshaped due to material displacement and gravitational settling. In most cases, the resulting surface approximates a linear slope defined by the material's angle of repose. Thus, it is common to model the stockpile surface as a sloped line in subsequent excavation cycles. However, in certain cases, such as in this simulation, the soil retains the shape carved during the initial excavation rather than collapsing into a uniform slope. This deviation from the linear-surface assumption can introduce errors in downstream computations, particularly when estimating the depth of penetration.

To address this concern, we modify the approach for computing penetration depth. Instead of assuming a triangular stockpile and measuring the depth as the distance $BD$ from the bucket tip to the idealized sloped surface (as shown in Fig. 2), we measure the distance directly to the actual excavated surface from Cycle 1. In the dual-cycle scenario illustrated in Fig. 9 this corresponds to measuring the depth relative to the red trajectory of the first cycle. This adjustment allows the model to better reflect the true geometry of the soil pile and improves force prediction accuracy in subsequent cycles.

In Cycle 2, the model achieves an RMSE of 17.3% by incorporating adaptive excavation depth estimation. By estimating depth relative to the actual excavated surface rather than assuming a fixed triangular wedge, the model better captures the true interaction geometry and improves force prediction accuracy. A more robust solution would involve an adaptive soil modeling framework in which the wedge geometry is no longer constrained to a constant stockpile angle or a fixed triangular form. Instead, it would dynamically conform to the actual shape of the evolving soil surface throughout excavation. Designing such a geometry-aware and adaptive soil–tool interaction model remains a promising direction for future work.

### C. SINGLE PATH DIFFERENT SOIL COMPARISON

To evaluate the adaptability of the proposed force estimation method across different soil types, several soil configurations in Algoryx [37] are tested, including gravel, sand, and dirt as summarized in Table 2. Due to the similarity between gravel and sand—both having identical density (1474 kg/m$^3$) and zero cohesion—a modified sand profile is introduced to represent clayey sand. This new profile uses an average density of 1874.5 kg/m$^3$ and a reduced internal friction angle of 0.59 rad, which is consistent with values reported for clayey sand in Table A1.

The resulting force trends obtained from parameter optimization for gravel, modified sand, and dirt are compared in Fig. 10. Table 4 compares estimated soil parameters with Algoryx reference values. For density, all estimated values (1513.4 for gravel, 1802.3 for modified sand, and 1833.2 kg/m$^3$ for dirt) fall within the range of 1297–2345 kg/m$^3$ as reported in the literature with deviations from Algoryx of 3.7%, 6.9%, and 34.3%, respectively.

For cohesion and adhesion, although the literature permits values up to 50000 N/m$^2$ for soft to firm cohesive soils, the Algoryx library uses a range of 0–2500 N/m$^2$ for the range of parameters presented. To maintain consistency with Algoryx, we adopt the 0–2500 N/m$^2$ bound when evaluating estimated results. The estimated cohesion values for gravel (143.8 N/m$^2$) and modified sand (216 N/m$^2$) are within this Algoryx-consistent range, while dirt returns $C = 0$ compared to the Algoryx value of 2100 N/m$^2$. This difference is compensated by higher adhesion (3039.5 N/m$^2$), which exceeds the 2500 N/m$^2$ Algoryx limit by 21.6% but remains in the range reported in the literature. Adhesion values for gravel (184 N/m$^2$) and modified sand (72.4 N/m$^2$) are well within range with a 7.36% and 2.9% error, respectively.

Internal friction angle estimates (0.78 rad for gravel and modified sand, 0.37 rad for dirt) fall within the literature range of 0.37–0.69 rad. The gravel and modified sand values slightly exceed Algoryx references (0.76 rad and 0.59 rad), while the dirt value matches the lower bound and remains physically consistent. Since ground truth values for the remaining parameters are not available from Algoryx, direct comparison is not possible.

It is important to note that the current optimization framework is designed to prioritize accurate force prediction by minimizing the discrepancy between observed and modeled forces using a streamlined cost function. This approach enables robust and data-efficient parameter fitting tailored for excavation force estimation. While the identified parameters are not constrained to match the physically measured soil values perfectly, they remain within realistic bounds and are effective for predictive modeling. Future extensions of this work may incorporate additional physical constraints or regularization techniques into the model to enhance parameter interpretability and further strengthen the connection between force modeling and soil characterization.

## V. CONCLUSION

This study presented a data-efficient framework for excavation force estimation in wheel loaders using hlreal-time soil parameter optimization and a modified analytical force model. The proposed method achieved an RMSE as low as 8.6% in force prediction and reduced computation time by over 45% through a multi-stage constrained optimization strategy.

By leveraging force data from a prior bucket loading cycle, the framework calibrated soil parameters, such as density, cohesion, adhesion, internal and external friction angles, and compaction, which were then used to predict resistive forces for the next cycle. This process enables adaptive, cycle-to-cycle excavation planning without requiring large datasets or offline training. The methodology integrates a modified Fundamental Earthmoving Equation with continuity constraints and Bekker's load–sinkage formulation, improving both numerical stability and physical relevance.

Validated through high-fidelity Algoryx Dynamics simulations, the approach can be effectively generalized across different trajectories and soil types. Its low data requirements and computational efficiency make it suitable for execution within the interval between digging cycles, supporting timely adaptation in autonomous excavation systems.

Future work will focus on extending the framework with dynamic force modeling to capture velocity, acceleration, and inertial effects as well as enhancing soil displacement and geometry representation during excavation, incorporating more systematic strategies for initialization sensitivity, and validating performance on experimental platforms beyond simulation. These efforts aim to address the current limitations of the quasi-static formulation, simplified stockpile assumptions, dependence on initialization, and reliance on simulation-based validation. In addition, future research will explore perception-based techniques, such as vision and depth sensing, to infer soil type and surface characteristics, thereby providing complementary information that can improve soil parameter accuracy.

## ACKNOWLEDGMENT

The authors acknowledge Algoryx Simulation, Sweden, for providing access to their simulation software. Additionally, they acknowledge the use of AI-based tools to help refine the clarity of the manuscript text. This work was funded in part by Komatsu Ltd. The authors gratefully acknowledge their financial support.

## APPENDIX A
## SOIL PARAMETER VALUES AND RANGES REPORTED IN LITERATURE

To establish realistic parameter bounds, values were compiled from multiple literature sources and engineering databases. StructX [41], an online platform for engineering data, provided soil phase relationships, densities, elastic modulus, Poisson's ratio, angles of repose, and properties of cohesive and non-cohesive soils. Soil density ranges were drawn from historical and contemporary references, including [42], while cohesive and frictional properties were compiled from handbooks such as [43]. Additional cohesion and internal friction angle values were obtained from EcorisQ [44], which reports soil parameters following the Unified Soil Classification System (USCS) [45]. Adhesion coefficient ranges were cross-referenced against GEO5 geotechnical software resources [46], a widely applied toolset in geotechnical practice.

The sources described above provide parameters such as soil density $\gamma$, cohesion $C$, adhesion $C_a$, and internal friction angle $\phi$ as functions of soil type or soil state, summarized in Table A1. The external friction angle $\delta$, which governs soil–structure interaction, depends on both the internal friction angle $\phi$ and the material properties of the contacting surface (e.g., steel, concrete, or timber). The Delft Sand, Clay, and Rock Cutting Model [47] reports typical relationships between $\delta$ and $\phi$, often expressing $\delta$ as a fraction of $\phi$. Representative values of external friction are included at the end of Table A1. For consistency with typical construction equipment, the wheel loader bucket is assumed to be made of steel.

In addition to the parameters summarized above, Bekker's load–sinkage formulation introduces three additional quantities: $k_c$ (cohesive modulus of deformation), $k_\phi$ (frictional modulus of deformation), and $n$ (exponent of deformation). These parameters characterize the non-linear pressure–sinkage relationship of soil under loading and are particularly relevant for modeling cutting and penetration resistance in excavation processes. Widely applied in terramechanics, Bekker's formulation enables the prediction of soil sinkage for wheels and tracks operating on soft, deformable terrains such as planetary surfaces, agricultural fields, and unpaved construction sites. Reported values for these parameters from multiple sources are compiled in Table A2.

**TABLE A1. Summary of soil parameters collected from multiple sources. Parameters are categorized based on either soil type (using the USCS classification), soil state or the material in contact with. Density $\gamma$, cohesion $C$, adhesion $C_a$, and internal friction angle $\phi$ are reported where available. External friction angle $\delta$ values are based on the material in contact with the soil, following relationships compiled from [47]. A dash (–) indicates that the value is not reported or not applicable for the given soil classification.**

| Soil Description | USCS Class | Density $\gamma$ (kg/m$^3$) | Cohesion $C$ (kPa) | Adhesion $C_a$ (kPa) | Internal Friction $\phi$° | External Friction $\delta$° |
|---|---|---|---|---|---|---|
| **Categorized by soil type. From [41], [42] and [43]** | | | | | | |
| Well-graded gravel, fine to coarse gravel | GW | 1631 – 1937 | 0 | – | 40 | – |
| Poorly graded gravel | GP | 1631 – 1937 | 0 | – | 38 | – |
| Silty gravel | GM | 1297 – 1500 | 0 | – | 36 | – |
| Clayey gravel | GC | 1297 – 1500 | 0 | – | 34 | – |
| Clayey gravel with fines | GC-CL | 1297 – 1500 | 3 | – | 29 | – |
| Well-graded sand, fine to coarse sand | SW | 1410 – 2279 | 0 | – | 38 | – |
| Poorly graded sand | SP | 1410 – 2279 | 0 | – | 36 | – |
| Silty sand | SM | 1378 – 2371 | 0 | – | 34 | – |
| Clayey sand | SC | 1378 – 2371 | 0 | – | 32 | – |
| Silt | ML | 1300 – 1380 | 0 | – | 33 | – |
| Clay of low plasticity, lean clay | CL | 1330 – 1390 | 20 | – | 27 | – |
| Clay of high plasticity, fat clay | CH | 1330 – 1470 | 25 | – | 22 | – |
| Organic silt, organic clay | OL | 1330 – 1500 | 10 | – | 25 | – |
| Organic clay, organic silt | OH | 1330 – 1500 | 10 | – | 22 | – |
| Silt of high plasticity, elastic silt | MH | 1300 – 1380 | 5 | – | 24 | – |
| **Categorized by soil state. From [44] and [46]** | | | | | | |
| Soft and very soft cohesive soil | – | – | 0 – 12 | 0 – 12 | – | – |
| Cohesive soil with medium consistency | – | – | 12 – 24 | 12 – 24 | – | – |
| Stiff cohesive soil | – | – | 24 – 48 | 24 – 48 | – | – |
| Hard cohesive soil | – | – | 48 – 96 | 48 – 96 | – | – |
| Very Soft Soil | – | 1631 – 1937 | 0 – 10 | - | – | – |
| Soft Soil | – | 1733 – 2039 | 10 – 25 | - | – | – |
| Firm Soil | – | 1784 – 2141 | 25 – 50 | - | – | – |
| Stiff Soil | – | 1835 – 2243 | 50 – 100 | - | – | – |
| Very Stiff Soil | – | 2141 – 2243 | 100 – 200 | - | – | – |
| Hard Soil | – | 2039 – 2345 | 200 – – | - | – | – |
| **External friction angle categorized by contact material from [47]** | | | | | | |
| Steel Piles (NAVFAC) | – | – | – | – | – | 20° |
| USACE | – | – | – | – | – | $0.67\phi$–$0.83\phi$ |
| Steel (Broms) | – | – | – | – | – | 20° |
| Concrete (Broms) | – | – | – | – | – | $3/4\phi$ |
| Timber (Broms) | – | – | – | – | – | $2/3\phi$ |
| Lindeburg | – | – | – | – | – | $2/3\phi$ |
| Concrete Walls (Coulomb) | – | – | – | – | – | $2/3\phi$ |

**TABLE A2. Pressure-Sinkage Parameters for Different Soil Types. Values have been gathered as reported in different literature. Table has been sectioned into values taken from different literature that has been cited.**

| Soil Type | $k_c$ (kN/m$^{n+1}$) | $k_\phi$ (kN/m$^{n+2}$) | n |
|---|---|---|---|
| **From simulation of the behavior of tracked vehicles on soft soils [48]** | | | |
| Dry Loose Sand | 0.00E+00 | 1.58E+03 | 1.01 |
| Dry Compact Sand | 9.57E+01 | 3.27E+03 | 1.15 |
| Dry Sand LLL[1] | 0.99E+00 | 1.52E+03 | 1.10 |
| Heavy Clay WES[2] 40 | 1.84E+00 | 1.03E+02 | 0.11 |
| Lean Clay WES[2] 32 | 1.52E+00 | 1.19E+02 | 0.15 |
| LETE[3] Sand | 1.02E+02 | 5.30E+03 | 0.79 |
| LETE[3] Sand 2nd | 6.94E+00 | 5.06E+02 | 0.71 |
| Sandy Loam | 1.19E+01 | 6.74E+02 | 0.81 |
| Soft Snow | 6.16E+00 | 1.49E+02 | 1.53 |
| IIT Dry Sand[4] [49] | -1.12E+02 | 3.101E+03 | 0.75 |
| **From carrier lunar parameter[5] [50]** | | | |
| Soil Type A | 0.00E+00 | 8.20E+02 | 1.00 |
| Soil Type B | 1.40E+00 | 8.20E+02 | 1.00 |
| Soil Type C | 2.80E+00 | 8.20E+02 | 1.00 |
| **From deformation of planetary soils [51]** | | | |
| Moon | 0.14E+00 | 8.20E+02 | 1.00 |
| Mars (MSS-A[6]) | 1.87E+01 | 7.63E+02 | 0.63 |
| Earth (Dry Sand) | 0.99E+00 | 1.52E+03 | 1.10 |
| Earth (Clay) | 1.31E+01 | 6.92E+02 | 0.50 |

[1] LLL: Lunar Logistics Load simulant soil.

[2] WES: Waterways Experiment Station soil classification.

[3] LETE: Land Engineering Test Establishment, Department of National Defence, Canada.

[4] IIT: Soil tested at Illinois Institute of Technology.

[5] Soil types A, B, and C refer to theoretical soil models developed during the lunar roving vehicle design phase to bracket expected lunar surface conditions. type A represents minimum soil strength properties, type B represents median (nominal) conditions, and Type C represents maximum soil strength properties.

[6] MSS-A: Mars Soil Simulant A.

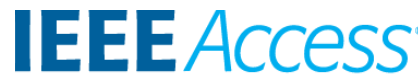

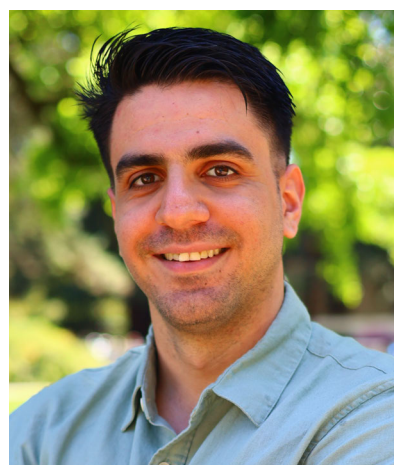

**ARMIN ABDOLMOHAMMADI** received the B.Sc. degree in mechanical engineering from the Sharif University of Technology, Tehran, Iran, in 2019, and the M.Sc. degree from the University of California at Davis, in 2025, where he is currently pursuing the Ph.D. degree with the Control, Optimization, Robotics, and Electrification (CORE) Laboratory. His research interests include system dynamics, control theory, and robotics, with a particular emphasis on optimal control and its applications in automotive and off-road machinery systems.

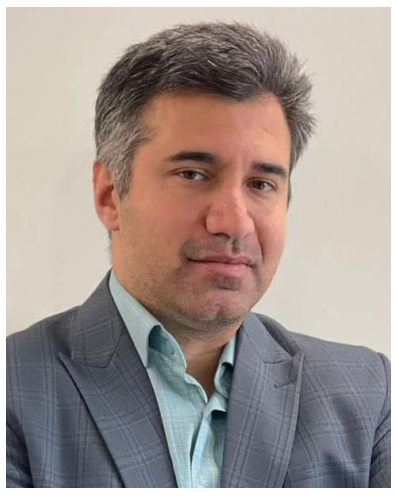

**NAVID MOJAHED** received the B.Sc. degree in mechanical engineering from the Isfahan University of Technology, Iran, the M.Sc. degree in applied mathematics from the Amirkabir University of Technology, Iran, and the Ph.D. degree in applied mathematics from the University of Mazandaran, Iran. He was a Researcher at the University of Santiago de Compostela, Spain. He is currently a Postdoctoral Scholar at the University of California at Davis. His research interests include fractional differential equations, numerical methods, game theory, optimal control, and autonomous vehicle systems.

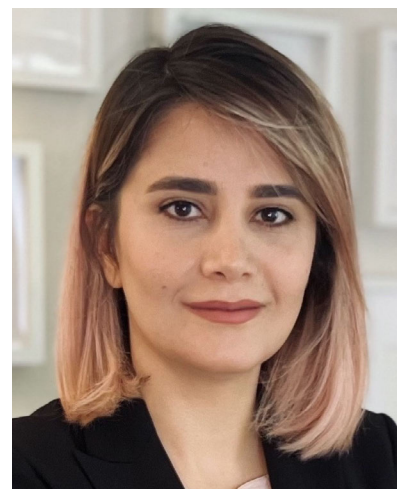

**SHIMA NAZARI** received the B.S. and M.S. degrees from the Sharif University of Technology, Tehran, Iran, in 2009 and 2012, respectively, and the Ph.D. degree in mechanical engineering from the University of Michigan, in 2019. She is currently an Assistant Professor of mechanical engineering with the University of California at Davis. Prior to this position, she was a Postdoctoral Researcher at the Model Predictive Control Laboratory, UC Berkeley. Her research interests include the areas of optimal and data-driven control with applications for automation, electric vehicles, and transportation systems.

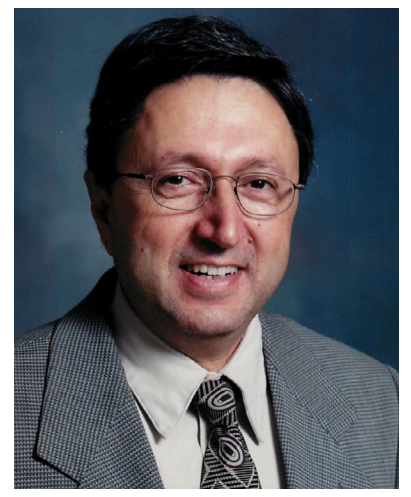

**BAHRAM RAVANI** received the B.S. degree (magna cum laude) in mechanical engineering from Louisiana State University, in 1976, the M.S. degree in mechanical engineering from Columbia University, in 1978, and the Ph.D. degree in mechanical engineering from Stanford University, in 1982.

He was a tenured Faculty Member at the University of Wisconsin–Madison. He has been a Faculty Member at UC Davis, since 1987, where he was the Chair of the Mechanical Engineering Department and the Electrical and Computer Engineering Department. He also served as the Campus Director of the Center for Information Technology Research in the Interest of Society. He was also the Founding Director of the Advanced Highway Maintenance and Construction Technology (AHMCT) Research Center, a collaboration between California Department of Transportation and UC Davis. He co-directed AHMCT, until July 2025. He is currently a Distinguished Professor of mechanical and aerospace engineering with UC Davis. His current research interests include the digital transformation of construction and maintenance, intelligent transportation systems and highway safety, kinematics, dynamics and biomechanics, robotics and mechatronics, mechanical design, and manufacturing.